\documentclass[conference,a4paper]{IEEEtran}

\usepackage{amssymb, amsthm, amsfonts}
\usepackage[utf8]{inputenc} 
\usepackage[T1]{fontenc}
\usepackage{url}  
\usepackage{tikz}
\usepackage[linesnumbered,algoruled,boxed,lined]{algorithm2e}

\usepackage[cmex10]{amsmath}  
\usepackage{mleftright}       
\mleftright                   

\usepackage{graphicx}         
\usepackage{booktabs}         

\usepackage{times}
\usepackage{graphicx}
\usepackage{setspace}
\usepackage{amsfonts}       
\usepackage{nicefrac} 
\usepackage{array}
\usepackage{bm}

\usepackage{tabularx}
\usepackage{balance}
\usepackage{hyperref}
\usepackage{mathtools}
\usepackage{nicematrix}
\newcommand{\etal}{\textit{et al. }}

\newcommand{\Fq}{\mathbb F_{q}}

\DeclareMathOperator{\res}{res}
\DeclareMathOperator{\sres}{sres}

\newcommand{\mat}[1]{\mathsf{#1}} 
 
\newcommand{\eqdef}{\triangleq} 
\newcommand{\Naturals}{\mathbb{N}}   
\newcommand{\Field}{\mathbb{F}}      
\newcommand*{\Scale}[2][4]{\scalebox{#1}{\ensuremath{#2}}} 

\usepackage[noadjust]{cite} 
\usepackage[capitalize]{cleveref} 
	\crefname{equation}{}{}
	\crefname{theorem}{Theorem}{Theorems}
	\crefname{lemma}{Lemma}{Lemmas}
	\crefname{cor}{Corollary}{Corollaries}
	\crefname{prop}{Proposition}{Propositions}
	\crefname{note}{Note}{Notes}
	\crefname{appsec}{Appendix}{Appendices}
	\crefname{definition}{Definition}{Definitions}
	\crefname{conj}{Conjecture}{Conjectures}
	\crefname{construction}{Construction}{Constructions}

\usepackage{enumitem}
\setlist[enumerate]{
	labelsep=8pt,
	labelindent=0pt,
	itemindent=0pt,
	leftmargin=*,
	before=\setlength{\listparindent}{-\leftmargin},
}

\newtheorem{theorem}{Theorem}
\newtheorem{example}{Example}%
\newtheorem{remark}{Remark}%
\newtheorem{lemma}{Lemma}%

\newtheorem{definition}{Definition}%
\usepackage{mathtools}

\theoremstyle{plain}
\theoremstyle{remark}

\newenvironment{continuance}[1]
  {\par\smallskip\noindent\textbf{Example #1 (Continued). }\itshape}
  {\par}
\allowdisplaybreaks

\newif\iflong
\longtrue 

\begin{document}

\title{Efficient Interpolation-Based Decoding of Reed-Solomon Codes} 


%
 \author{%
   \IEEEauthorblockN{Wrya K. Kadir, Hsuan-Yin Lin, and Eirik Rosnes}
   \IEEEauthorblockA{Simula UiB, 
                     N--5006 Bergen, Norway\\
                     Emails:\{wrya, lin, eirikrosnes\}@simula.no}
                  }

\maketitle


\begin{abstract}
We propose a new interpolation-based error decoding algorithm for $(n,k)$ Reed-Solomon (RS) codes over a finite field of size $q$, where $n=q-1$ is the length and $k$ is the dimension. In particular, we employ the fast Fourier transform (FFT) together with properties of a circulant matrix associated with the error interpolation polynomial and some known results from elimination theory in the decoding process. The asymptotic computational complexity of the proposed algorithm for correcting any $t \leq \lfloor \frac{n-k}{2} \rfloor$ errors in an $(n,k)$ RS code is of order $\mathcal{O}(t\log^2 t)$ and $\mathcal{O}(n\log^2 n \log\log n)$ over FFT-friendly and arbitrary finite fields, respectively, 
 achieving the best currently known asymptotic decoding complexity, proposed for the same set of parameters.
\end{abstract}

\section{Introduction}
Reed-Solomon (RS) codes are the most widely known family of maximum distance separable (MDS) codes and were introduced by Reed and Solomon~\cite{reed1960polynomial}. 
RS codes have been used in secret sharing~\cite{mceliece1981sharing}, space communications, consumer electronics~\cite{wicker1999reed}, and QR codes~\cite{soon2008qr}. Hence, it is of great interest to devise efficient error decoding algorithms for RS codes.

For an $(n,k)$ RS code, where $n$ is the code length and $k$ is the message length (dimension), there are two main hard-decision decoding approaches, namely, syndrome-based and interpolation-based decoding. Berlekamp and Massey~\cite{massey1969shift} introduced a decoding algorithm in 1969, and later Sugiyama \etal \cite{sugiyama1975method} used the Euclidean algorithm to solve the \emph{key equation} of this algorithm. Both algorithms are syndrome-based decoding algorithms with computation complexity $\mathcal{O}(n(n-k))$~\cite{blahut2003algebraic,moon2020error,decode-newton-interpol93}. Welch and Berlekamp~\cite{welch1986error} introduced the first interpolation-based RS decoder 
with complexity $\mathcal{O}(n^3)$ (see also \cite[Th.~17.1.4]{guruswami2012essential}).

During the last two decades, several new decoding algorithms have been proposed for RS codes under the light of the (multiplicative) fast Fourier transform (FFT). Justesen in~\cite{justesen1976complexity} used the FFT instead of the method by Sugiyama \etal to solve the key equation and achieved the complexity $\mathcal{O}(q\log^2q)$, where $q\in \bigl\{2^{2^i}+1|\,i=0,1,2,3, 4\bigr\}$ is the field size.
The interpolation-based decoding algorithm proposed by Gao~\cite{gao2003new} for codes with length $n$ over an \emph{FFT-friendly} finite field \cite{von2013modern} of size $q$, where $q-1$ is divisible by $n$, has complexity $\mathcal{O}(n\log^2 n)$. 

For RS codes over fields with even characteristics, an additive FFT~\cite{gao2010additive,wang1988fast,cantor1989arithmetical} can be applied. A comparison study between the syndrome-based algorithms~\cite{massey1969shift,moon2020error,sugiyama1975method} and the interpolation-based decoding algorithm~\cite{gao2003new} was done in~\cite{chen2008complexity}. The study shows that the interpolation-based algorithm~\cite{gao2003new} is more efficient than syndrome-based decoders only for RS codes of very low rate.  New versions of additive FFTs were introduced in \cite{S.j.lin2016-1,tang2022new}, and additive FFT-based decoding algorithms on RS codes appeared in~\cite{tang2022new,lin2016fft}. These algorithms require finite fields with even characteristics and additional constraints, such as $n$, $k$, or $n-k$ being a power of two.
The new additive FFTs are not compatible with our code parameters. 

\begin{table}[t!]
  \centering
  {\scriptsize
    \caption{Complexity Comparison Between our Algorithm and Algorithms in \cite{sugiyama1975method,massey1969shift,justesen1976complexity,gao2003new,tang2022new,lin2016fft,chen2008complexity,welch1986error,guruswami2012essential}}
        \vspace{-2ex}
    \begin{NiceTabular}{l@{\hspace{7pt}}l@{\hspace{7pt}}l}
    \toprule
         Decoding Methods & Complexity & Restriction\tabularnote{$\Fq$ denotes a finite field of size $q$, $a | b$ denotes  that $a$ is a divisor of $b$, and $\Naturals$ represents the set of natural numbers.}\\
          \midrule
          Syn.-based \cite{massey1969shift,sugiyama1975method}& $\mathcal{O}(n(n-k))$ & arb. $\Fq$\\
         Int.-based \cite{welch1986error,guruswami2012essential} & $\mathcal{O}(n^3)$ &arb. $\Fq$\\
        Syn.-based \cite{justesen1976complexity} & $\mathcal{O}(q\log^2 q)$ & $q\in \bigl\{2^{2^i}+1|\, 0 \leq i \leq 4\bigr\}$ \\[1mm]
           Syn.-based \cite{lin2016fft} & $\mathcal{O}((n-k)\log^2 (n-k))$ & $n=q=2^m$, $n-k=2^\lambda$, \\
           & & $m,\lambda\in\Naturals$, and $\nicefrac{k}{n}\geq \nicefrac{1}{2}$ \\
          Syn.-based \cite{tang2022new} & $\mathcal{O}((n-k)\log^2 (n-k))$ & $n=q=2^m$, $m\in\Naturals$ \\
        Int.-based \cite{gao2003new} & $\mathcal{O}(n\log^2 n)$  & FFT-fri. $\Fq$, $n|(q-1)$ \\
          Our method & $\mathcal{O}(t\log^2 t)$\tabularnote {$t$ is of order $\mathcal{O}(n)$, but smaller.}  & FFT-fri. $\Fq$, $n=q-1$ \\
           Int.-based \cite{gao2003new,chen2008complexity} & $\mathcal{O}(n\log^2 n\log \log n)$  & $\Fq$ with char. $2$, $n|(q-1)$ \\
            Our method & $\mathcal{O}(n\log^2 n \log\log n)$  & arb. $\Fq$, $n=q-1$ \\
             \bottomrule
    \end{NiceTabular}
    \label{tab:table-complex-com}}
    \vspace{-5ex}
\end{table}

In this paper, we develop a new interpolation-based error decoding algorithm for $(n,k)$ RS codes over finite fields of size $q$ and for any $t \leq \lfloor \frac{n-k}{2} \rfloor$ errors with block length $n=q-1$. Our algorithm is based on a simple observation, yielding the decoding complexity $\mathcal{O}(t\log^2 t)$ over an FFT-friendly finite field and $\mathcal{O}(n\log^2 n\log\log n)$  over an arbitrary finite field (see Theorem \ref{Th:main-theorem}).
The interpolation-based algorithm in \cite{gao2003new} can also achieve the same complexity order, but only over finite fields with even characteristic.
We use properties of the error interpolation polynomial (see \cref{Lemma:poly_g(v=n-t)} and \cref{Th:main}) and its associated circulant matrix in the decoding process. Finding the error locator polynomial and its roots are the most time-consuming steps in syndrome-based algorithms while finding the greatest common divisor (GCD) between two polynomials is the step with the highest time complexity for the existing interpolation-based algorithms, but none of these steps are required for our algorithm. As an alternate, we reduce the decoding problem to the problem of solving a Toeplitz linear system of equations which can be solved with the same complexity as finding a GCD \cite{brent1980fast}.  
The complexity analysis in Section~\ref{sec:com-analysis} shows the asymptotic complexity of the proposed algorithm, and   Table \ref{tab:table-complex-com} compares it to algorithms in \cite{massey1969shift,sugiyama1975method,welch1986error,justesen1976complexity,gao2003new,tang2022new,lin2016fft,chen2008complexity,guruswami2012essential}.
  A similar idea
  that uses the rank properties of  Dickson matrices has been used in decoding rank-metric codes \cite{kadir2020decoding,kadir2021interpolation,kadir2022decoding,kadir2021new}.

 \section{Notation}
 
Let $\mathbb{F}_q$ denote a finite field with $q$ elements and  $\mathbb{F}_q[x]$  a univariate polynomial ring over $\Fq$. The degree of a polynomial $f(x) \in \mathbb{F}_q[x]$ is the largest of the degrees of the individual terms and is denoted by $\deg(f(x))$. For simplicity, we sometimes write $f$ instead of $f(x)$  if the dependency on $x$ is clear from the context, and we also use $\bm{f}$ to denote the vector of coefficients of $f(x)$. We denote the GCD of two polynomials $f,g$ by $\gcd(f,g)$.  Let $\alpha$ be a primitive $n$-th root of unity, i.e., $n$ is the smallest positive integer such that $\alpha^n=1$. The evaluation of a polynomial of degree less than $n$ on distinct points in  $\mathcal{A}\triangleq \{\alpha^0,\ldots, \alpha^{n-1}\}$ is referred to as the discrete Fourier transform (DFT). The interpolation is the inverse transform, called the inverse discrete Fourier transform (IDFT). FFT refers to an algorithm that computes the DFT of length $n$ in time complexity $\mathcal{O}(n(\log n)^u)$, for some small $u$. A (multiplicative) FFT of length $n$ requires that $n$ is the product of only small prime numbers and also that $x^n-1$ has $n$ distinct roots in $\mathbb{F}_q$. Since we consider $n=q-1$, the second condition always holds. A finite field that provides these two properties is referred to as FFT-friendly~\cite{gao2003new,von2013modern}.

Let $\rho_{-1},\rho_0\in \mathbb{F}_q[x]$ and $\rho_0\ne 0$. The \textit{Euclidean remainder sequence}  is defined as 
$\rho_{i-2}=q_i\rho_{i-1}+\rho_{i}$, $\deg(\rho_{i})<\deg(\rho_{i-1})$, for $1\leq i\leq s$, where  $\rho_i,q_i\in\Field_q[x]$ are the $i$-th remainder and quotient, respectively, and $s$ is the smallest positive integer for which $\rho_{s}\mid\rho_{s-1}$. $\rho_s=\gcd(\rho_{-1},\rho_0)$.

\section{Decoding Reed-Solomon Codes}
An $(n,k)$  RS code $\mathcal{C}$ with length $n$, dimension $k$, and minimum distance $n-k+1$ over $\Fq$ is a classical example of a polynomial evaluation MDS code. A message vector $\bm{f} =(f_0,\ldots, f_{k-1})\in \Fq^k$ is treated as the coefficient vector of a polynomial of degree less than $k$ and evaluated over $\mathcal{A}$. The output is an RS codeword. 

\subsection{Encoding and Error Generation}
The message $\bm{f}=(f_0,f_1,\ldots,f_{k-1})\in \mathbb{F}_q^k$, with corresponding message polynomial $f(x)=f_0+f_1x+\cdots+f_{k-1}x^{k-1}\in \mathbb{F}_q[x]$, is padded by $n-k$ zeros and encoded as 
$\bm{c}=\tilde{\bm{f}}\cdot \mat{G}_{\alpha}$, 
where 
$\mat{G}_\alpha=(\mat{G}_{\alpha})_{i,j}=\alpha^{j(i-1)} \mbox{, for } 0\leq i,j\leq n-1$, 
is a \emph{generator} matrix of $\mathcal{C}$, $\tilde{\bm{f}}\eqdef(f_0\ldots,f_{k-1},0\ldots,0)\in \Fq^n$,  
$\alpha$ is an $n$-th primitive root of unity, and $n=q-1$. 
Hence, encoding RS codes is equivalent to computing a DFT. The matrix $\mat{G}_{\alpha}$ is a nonsingular \textit{Vandermonde} matrix \cite{lidl1997finite}.

The channel chooses a random vector $\bm{e}\in \mathbb{F}_q^n$ of weight $t$ where $t\leq \lfloor\frac{n-k}{2}\rfloor$. Let $\Tilde{g}(x)\in \mathbb{F}_q[x]$ be the error interpolation polynomial with coefficient vector $\tilde{\bm{g}}=(g_0,\ldots, g_{n-1})\in \mathbb{F}_q^n$ such that 
$\bm{e}=\tilde{\bm{g}}\cdot \mat{G}_{\alpha}.$
The error vector $\bm{e}$ is added to our codeword $\bm{c}$ and we get  
\begin{equation}\label{eq:r=(f+g)G}
    \bm{r}=\bm{c}+\bm{e}=\tilde{\bm{f}}\cdot \mat{G}_{\alpha}+\tilde{\bm{g}}\cdot \mat{G}_{\alpha}=(\tilde{\bm{f}}+\tilde{\bm{g}})\cdot \mat{G}_{\alpha},
\end{equation}
where $\bm{r}$ denotes the received vector. 
\subsection{Decoding}
\label{sec:decoding}
\subsubsection{Circulant Matrix}
We recall the following known result from \cite{von2013modern}. 
\begin{lemma}\label{lem:G&G^-1}
Let $\mathbb{R}$ be a commutative ring with identity, $n$ a positive integer, and $\alpha\in \mathbb{R}$ a primitive $n$-th root of unity. Then, $\alpha^{-1}$ is also a primitive $n$-th root of unity and $\mat{G}_{\alpha}\mat{G}_{\alpha^{-1}}=n\mat{I}$, where $\mat{I}$ is the identity matrix of order $n$.
\end{lemma}
Using Lemma \ref{lem:G&G^-1}, computing $\mat{G}_{\alpha}^{-1}=\frac{1}{n}\mat{G}_{\alpha^{-1}}$ is easy and can be done in advance. Multiplying a vector by $\mat{G}_{\alpha}^{-1}$ is an IDFT.
 Matrix $\mat{G}_{\alpha}$ is nonsingular and we can multiply both sides of \eqref{eq:r=(f+g)G} by $\mat{G}_{\alpha}^{-1}$ and get 
 \begin{equation}   
   \bm{r}\cdot \mat{G}_{\alpha}^{-1}=(\tilde{\bm{f}}+\tilde{\bm{g}})\cdot \mat{G}_{\alpha}\cdot \mat{G}_{\alpha}^{-1}=\tilde{\bm{f}}+\tilde{\bm{g}}.\label{eq:rG^-1=f+g}
 \end{equation}
Let $\bm{\beta}=(\beta_0,\ldots, \beta_{n-1})=\bm{r}\cdot \mat{G}_{\alpha}^{-1}$. The decoder knows $\bm{r}$ and $\mat{G}_{\alpha}^{-1}$, and hence $\bm{\beta}$. One can write \eqref{eq:rG^-1=f+g} as 
\begin{IEEEeqnarray}{rCl}
  (\beta_0,\ldots,\beta_{n-1})& = &(f_0,\ldots,f_{k-1},0,\ldots,0)
  \nonumber\\
  && +\>(g_0,\ldots,g_{k-1},g_k,\ldots, g_{n-1}).
  \IEEEeqnarraynumspace\label{eq:beta=f+g}
\end{IEEEeqnarray}
The decoder needs to recover $\Tilde{g}(x)$ and evaluate $\Tilde{g}(x)$ on $\alpha^0,\dotsc,\alpha^{n-1}$ to get the error values. It can be seen from \eqref{eq:beta=f+g} that  the $n-k$ coefficients $g_k,\ldots,g_{n-1}$ of $\Tilde{g}(x)$ are known and the remaining task is to find the $k$ coefficients $g_0,\ldots,g_{k-1}$.
\begin{example}\label{example1}
Let $\mathcal{C}$ be an RS code over $\mathbb{F}_{11}$ with $n=10$, $k=4$, and $\alpha=2\in \mathbb{F}_{11}$. Let $\bm{r}=( 8, 0, 4, 3, 6, 10, 1, 8, 4, 3 )\in \mathbb{F}_{11}^{10}$ be a received vector via a noisy channel with $t=3$ errors, and 
let $\mat{G}_{\alpha}$ be the Vandermonde matrix associated with $\alpha$. Then, $\bm{\beta}=\bm{r}\cdot \mat{G}_{\alpha}^{-1}=(8, 0, 9, 0, 2, 1, 8, 7, 4, 2 )$.
Since $\bm{\beta}=\tilde{\bm{f}}+\tilde{\bm{g}}$, we already know the last $n-k=6$ coefficients of $\tilde{\bm{g}}$ and $\beta_i=g_i$ for $4\leq i\leq 9$, i.e., $(g_4,\ldots, g_{9})=(2,1,8,7,4,2)$.

\end{example}
\begin{definition}
Let $\bm{a}=(a_0,a_1,\ldots, a_{d-1})\in \mathbb{F}_q^d$. The matrix of the form 
  \begin{equation}\label{eq:circulantmat}
  \mat{A}= \left(
    \begin{array}{cccccc}
      a_0&a_1&\cdots&a_{d-3}&a_{d-2}&a_{d-1}
      \\
      a_{d-1}&a_0&\cdots&a_{d-4}&a_{d-3}&a_{d-2}
      \\
      a_{d-2}&a_{d-1}&\cdots&a_{d-5}&a_{d-4}&a_{d-3}
      \\
      \vdots&\vdots&\ddots&\vdots&\vdots&\vdots
      \\
      a_{1}&a_{2}&\cdots&a_{d-2}&a_{d-1}&a_{0}
    \end{array}
  \right)
\end{equation}%
is the \textbf{circulant matrix} associated with vector $\bm{a}$, where each row is the right circular shift of the row above.
\end{definition}
A circulant matrix is a special case of a Toeplitz matrix and  all the square submatrices of a circulant matrix have a Toeplitz structure as all the entries along the diagonal are equal, and those along each line parallel to the diagonal are also equal. A $u\times u$ Toeplitz matrix can be  determined by its $2u-1$,  $u\in\Naturals$,  entries in its first row and first column.
\begin{theorem}[K\"{o}nig-Rado's theorem~\cite{lidl1997finite}\footnote{The formulation in \cite{lidl1997finite} is for the ``reverse'' direction, but the proof holds for both directions.}]\label{Th:konig-rados}
Let $a(x)=a_0+a_1x+\cdots+a_{q-2}x^{q-2}\in \mathbb{F}_q[x]$ be a polynomial of degree less than $q-1$ and let $\bm{a}=(a_0,\ldots, a_{q-2})$ be its coefficient vector. Then, the circulant matrix associated with $\bm{a}$  of the form in \eqref{eq:circulantmat}, 
has rank $\tau$ if the number of nonzero roots of $a(x)$ is $q-1-\tau$. 
\end{theorem}

We write the circulant matrix associated with the coefficient vector $\tilde{\bm{g}}=(g_0,\ldots,g_{n-1})$  as
\setlength{\arraycolsep}{2pt}
\begin{IEEEeqnarray}{c} \label{eq:tildeG}
  \label{eq:g_circulantmat}
  \tilde{\mat{G}}=
  \begin{pNiceArray}{ccc|cccc}
    g_0&g_1&\cdots&\textcolor{blue}{g_{n-t-1}}&\textcolor{red}{g_{n-t}}&\textcolor{red}{\ldots}&\textcolor{red}{g_{n-1}} \\
    g_{n-1}&g_0&\cdots&\textcolor{blue}{g_{n-t-2}}&\textcolor{red}{g_{n-t-1}}&\textcolor{red}{\ldots}&\textcolor{red}{g_{n-2}}  \\
    \vdots&\vdots&\cdots&\textcolor{blue}{\vdots}&\textcolor{red}{\vdots}&\textcolor{red}{\ldots}&\textcolor{red}{\vdots}  \\
    g_{n-t+1}&g_{n-t+2}&\cdots&\textcolor{blue}{g_{n-2t}}&\textcolor{red}{g_{n-2t+1}}&\textcolor{red}{\ldots}&\textcolor{red}{g_{n-t}} \\
    \hline
    \vdots&\vdots&\ddots&\vdots&\vdots&\ddots&\vdots\\
    g_{1}&g_{2}&\cdots&g_{n-t}&g_{n-t+1}&\ldots&g_{0} \\
  \end{pNiceArray},~
\end{IEEEeqnarray}
where $t$ is the number of errors. Hence,  $\bm{e}$ has $n-t$ zero components which correspond to  the number of nonzero distinct roots of $\Tilde{g}(x)$. Using Theorem~\ref{Th:konig-rados} one can conclude that the rank of $\tilde{\mat{G}}$ is $t$, and therefore it contains a $t\times t$ full rank submatrix. We are going to show that the Toeplitz $t\times t$ submatrix formed by the first $t$ rows and the last $t$ columns of $\tilde{\mat{G}}$ (colored in red) is always nonsingular. 
All the entries located in the top right $t\times (t+1)$ submatrix of $\tilde{\mat{G}}$ are known and we employ them to find the unknown coefficients $g_0,\ldots,g_{k-1}$. 

\begin{continuance}{\ref{example1}}
  The circulant matrix $\tilde{\mat{G}}$ associated with $\tilde{\bm{g}}$ has the form
  \begin{IEEEeqnarray}{c} \label{eq:tildeGexample}
    \Scale[1.0]{\mat{\Tilde{G}}=\begin{pNiceArray}{cccccc|cccc}          
 g_{0}&  g_{1} & g_{2}&  g_{3}&  2 & 1 & \textcolor{blue}{8} & \textcolor{red}{7}&  \textcolor{red}{4} & \textcolor{red}{2}\\
 2&  g_{0}&  g_{1} & g_{2}&  g_{3}&  2&  \textcolor{blue}{1}&  \textcolor{red}{8}&  \textcolor{red}{7}&  \textcolor{red}{4}\\
 4&  2 & g_{0}&  g_{1} & g_{2}&  g_{3}&  \textcolor{blue}{2}&  \textcolor{red}{1}&  \textcolor{red}{8}&  \textcolor{red}{7}\\
 \hline
 7&  4&  2&  g_{0}&  g_{1} & g_{2}&  g_{3} & 2&  1&  8\\
 8&  7&  4 & 2 & g_{0}&  g_{1} & g_{2}&  g_{3} & 2 & 1\\
 1&  8 & 7 & 4 & 2 & g_{0}&  g_{1} & g_{2}&  g_{3} & 2\\
 2&  1&  8 & 7 & 4 & 2 & g_{0}&  g_{1} & g_{2}&  g_{3}\\
 \hline
 g_{3} & 2 & 1 & 8 & 7 & 4 & 2 & g_{0}&  g_{1} & g_{2}\\
 g_{2}&  g_{3}&  2&  1&  8&  7&  4&  2& g_{0}&  g_{1}\\
 g_{1}&  g_{2}&  g_{3} & 2 & 1 & 8 & 7 & 4 & 2&  g_{0}
    \end{pNiceArray}},
  \end{IEEEeqnarray}
  and using Theorem \ref{Th:konig-rados}, its rank is $t=3$.
\end{continuance}

\subsubsection{Subresultants}

\begin{definition}[\hspace{-0.1ex}{\cite{von2013modern}}]
  Let $b(x)=\sum\limits_{i=0}^{d}b_ix^i$ and $a(x)=\sum\limits_{j=0}^{m}a_jx^j$ be two polynomials of degree $d$ and $m$ in $\mathbb{F}_q[x]$, respectively, where $d\geq m$. The matrix of the form 
  \setlength{\arraycolsep}{2pt}
  \begin{IEEEeqnarray*}{c}
    \mat{S}(b,a)\eqdef
      \Scale[0.950]{ \begin{pNiceArray}{cccc|ccccc}[cell-space-limits=0.25pt]
      b_{d}  &       &      &       & a_{m}  &       &      &      &
      \\            
      b_{d-1}&b_{d}  &      &       & a_{m-1}& a_{m} &      &      &
      \\      
      \Vdots &b_{d-1}&\ddots&       &\Vdots  &a_{m-1}&\ddots&      &
      \\      
      &\Vdots &\ddots&b_{d}  &a_1     &\Vdots &\ddots&      &
      \\
             &       &      &b_{d-1}&a_0     &a_1    &      &      &a_{m}
      \\
      b_0    &       &      &\Vdots &        &a_0    &\ddots&      &a_{m-1}
      \\     
             & b_0   &      &       &        &       &\ddots&      &\Vdots 
      \\        
             &       &\ddots&       &        &       &      &    \ddots  &a_1
      \\
      &       &      &b_0    &        &       &      &      &a_0\\
      \CodeAfter      
      \UnderBrace[shorten,yshift=3pt]{1-1}{9-4}{m \textnormal{ columns}}
      \UnderBrace[shorten,yshift=3pt]{1-5}{9-9}{d \textnormal{ columns}}
    \end{pNiceArray}}\IEEEeqnarraynumspace\\[3mm]
  \end{IEEEeqnarray*}
\normalsize
is called the \textbf{Sylvester matrix} of polynomials $b$ and $a$. The first $m$ columns are equipped by $b_i$'s and the last $d$ columns by $a_j$'s. All the entries outside of the two parallelograms are equal to zero. The \textbf{resultant} of $b$ and $a$ is the determinant of the Sylvester matrix and we denote it by $\res(b,a) \in \Fq$.
\end{definition}
\begin{definition}[\hspace{-0.1ex}\cite{von2013modern}]
For $0\leq l\leq m$, the determinant  of the $(m+d-2l)\times (m+d-2l)$ Sylvester submatrix 
 \setlength{\arraycolsep}{2.5pt}
  \begin{IEEEeqnarray*}{c}
    \mat{S}_l(b,a)\eqdef
     \Scale[0.950]{ \begin{pNiceArray}{cccc|ccccc}
      b_{d}      &      &      &      & a_{m}     &       &      &      &
      \\            
      b_{d-1}    &b_{d} &      &      & a_{m-1}   & a_{m} &      &      &
      \\      
      \Vdots     &      &\Ddots&      &\Vdots     &       &      &      &
      \\      
      b_{d-m+l+1}&\cdots&\cdots&b_{d} &a_{l+1}    &\Ddots &      &      &
      \\
      \Vdots     &      &      &\Vdots&\Vdots     &       &      &      &
      \\
      b_{l+1}    &\Cdots&      &b_{m} &a_{m-d+l+1}&\Cdots &      &      &a_{m}
      \\     
      \Vdots     &      &      &\Vdots&\Vdots     &       &      &      &\Vdots 
      \\        
      b_{2l-m+1} &\Cdots&      &b_l   &a_{2l-d+1} &\Cdots &      &      &a_l\\
      \CodeAfter      
      \UnderBrace[shorten,yshift=3pt]{1-1}{8-4}{m-l \textnormal{ columns}}
      \UnderBrace[shorten,yshift=3pt]{1-5}{8-9}{d-l \textnormal{ columns}}
    \end{pNiceArray}}\IEEEeqnarraynumspace\\[5mm]
  \end{IEEEeqnarray*}
\normalsize
is called the \textbf{$l$-th subresultant} of polynomials $b$ and $a$ and it is denoted by $\sres_l(b,a)\in \mathbb{F}_q$. The first $m-l$ columns of $\mat{S}_l(b,a)$ are equipped by $b_i$'s and the rest by $a_j$'s. By convention, a $b_i$ and an $a_j$ are zero for $i,j<0$. The $0$-th subresultant is the resultant of $b$ and $a$. 
\end{definition}
\subsubsection{Relation Between $\tilde{\mat{G}}$ and $\mat{S}(b,a)$}
 We review an important result from \cite{von2013modern} (Theorem~\ref{thm:Euclidean_rem}) and present a new key theorem (Theorem~\ref{Th:main}). 

\begin{theorem}[\hspace{-0.1ex}\cite{von2013modern}] \label{thm:Euclidean_rem}
Consider two polynomials $b,a\in \Fq[x]$  with degrees  $d\geq m$, respectively, and let $0\leq l\leq  m$. A polynomial of degree $l$ appears in the Euclidean remainder sequence of $b$ and $a$ if and only if $\sres_l(b,a)\neq 0$. 
\end{theorem}
The $l$-th subresultant of $b$ and $a$ is the leading coefficient of a polynomial of degree $l$ appearing in the Euclidean remainder sequence, and being nonzero means such a polynomial exists. 
\begin{theorem}\label{Th:main}
Let $b(x)=x^{q-1}-1$ and $a(x)=\sum\limits_{j=0}^{m}a_jx^j$  be two polynomials in $\mathbb{F}_q[x]$, where $a(x)=h(x)\cdot z(x)$ and $h(x)=\sum\limits_{i=0}^{v}h_ix^i$ has exactly $v\leq m\leq q-1$ distinct nonzero simple roots. The polynomial $a(x)=h\cdot z=h\cdot s\cdot r\cdot x^u$ satisfies one of the following:
\begin{itemize}
    \item[(i)] $s,r\in \mathbb{F}_q$ and $0\leq u\leq m-v$.  
    \item[(ii)] $s\in \Fq$, $r\in \Fq[x]$, $0\leq u\leq m-v-\deg(r)$, and factors of $r$ appear in the factors of $h$ if $\deg(r)\neq 0$. 
    \item[(iii)] $r,s\in \Fq[x]$, $0\leq u\leq m-v-\deg(r)-\deg(s)$, and factors of $r$ appear in the factors of $h$ if $\deg(r)\neq 0$. 
 Polynomial $s$ is irreducible over $\mathbb{F}_q[x]$ with a degree of at least two.
\end{itemize}
    
 Then, $\gcd(b,a)=h$ and  $\sres_v(b,a)\neq 0$.
\end{theorem}
\begin{IEEEproof}
\begin{itemize}
\item[(i)] If $a$ satisfies (i) and $u=0$, then $a=r\cdot s\cdot h$ and hence $b$ and $a$ share $v$ distinct linear factors and $\gcd(b,a)=h$. Therefore, a degree  $v$ polynomial appears in the remainder sequence and  $\sres_v(b,a)\neq 0$.

If $a$ satisfies (i) and $u\neq 0$, then $a=r\cdot s\cdot  h\cdot x^u$. Since $b$ does not have zero as a root, $\gcd(b,x^u)=1$, and again $b$ and $a$ share $v$ distinct linear factors and $\gcd(b,a)=\gcd(b,h)=h$. It is easy to see that factors of $x^u$ do not contribute in $\gcd(b,a)$ so we do not consider them in our analysis for the remaining cases (ii) and (iii). 
\item[(ii)] If $a$ satisfies (ii) and $r$ has $p\leq \deg(r)$ distinct nonzero roots, then $\gcd(h,r)$ is a degree $p$ polynomial which is the multiplication of $r$'s  distinct linear factors which are also factors of $h$. So again $\gcd(b,a)=\gcd(b,h)=h$. 
\item[(iii)] If $a$ satisfies (iii), then $s$ is irreducible and it is pairwise coprime with $b$, $h$, $r$, and $h\cdot r$, and also $s\cdot x^u$ is coprime with $b$. Since $\gcd(b,r\cdot x^u)=r$, by properties of GCDs of two polynomials, we have $\gcd(b,h\cdot r\cdot x^u\cdot s)=\gcd(b,h)=h$. 
\end{itemize}

In all cases, we have $\gcd(b,a)=h$ and  there exists a polynomial with degree $v$ in the remainder sequence. Finally, using Theorem~\ref{thm:Euclidean_rem} we conclude that $\sres_v(b,a)\neq 0$.
\end{IEEEproof}
\begin{remark} 
Note that $\sres_v(x^{q-1}-1,a)$ 
  in Theorem~\ref{Th:main} is the determinant of the matrix $\mat{S}_v(x^{q-1}-1,a)=$ 
  \setlength{\arraycolsep}{2pt}
  \begin{IEEEeqnarray}{c} \label{eq:Sv}
    \Scale[0.950]{
    \begin{pNiceArray}{cccc|cccccc}
      1&~&~&~ &a_{m}&~&~&~ && \\ 
      0&1&~&~&a_{m-1}&a_{m}&~&~&&\\ 
      \vdots&\vdots&\ddots&~&\vdots&~&\ddots&~&&\\
      0&\vdots&\ddots&1&a_{v+1}&\ldots&\ldots&a_m&~&\\
            0&\vdots&\ddots&0&\textcolor{red}{a_{v}}&\textcolor{red}{\ldots}&\textcolor{red}{\ldots}&~&\textcolor{red}{a_m}&\\
     \vdots&\vdots&\ddots&\cdots&\textcolor{red}{\vdots}&\textcolor{red}{\ddots}&\textcolor{red}{\ddots}&~&\textcolor{red}{\ddots}&\\
      0&\vdots&\ddots&0&\textcolor{red}{a_{m-q+v+2}}&\textcolor{red}{\ldots}&\textcolor{red}{\ldots}&\textcolor{red}{\ldots}&\textcolor{red}{\ldots}&\textcolor{red}{a_m}\\
      \vdots&~&~&\vdots&\textcolor{red}{\vdots}&~&~&~&~&\textcolor{red}{\vdots}\\
      0&\ldots&\ldots&0&\textcolor{red}{a_{2v-q+2}}&\textcolor{red}{\ldots}&\textcolor{red}{\ldots}&\textcolor{red}{\ldots}&\textcolor{red}{\ldots}&\textcolor{red}{a_v} \\    \end{pNiceArray},}\IEEEeqnarraynumspace\label{eq:matrix:Sv(f,g)}
  \end{IEEEeqnarray}
where the determinant of $\mat{S}_v(x^{q-1}-1,a)$ is equal to the determinant of its $(q-1-v) \times (q-1-v)$ Toeplitz submatrix (colored in red) in the downright corner with entries $a_{2v-q+2},\ldots,a_{v-1},a_v,a_{v+1}\ldots,a_m,0,\ldots,0$  and $a_v$ on the main diagonal. 
  In other words, if polynomial $a$ satisfies the properties in Theorem~\ref{Th:main}, then the Toeplitz submatrix in $\mat{S}_v(x^{q-1}-1,a)$ is always nonsingular. This observation is the key point for the proposed decoding algorithm. 
%
%
%
%
  %
%
%
%
   %
\end{remark}
\begin{lemma}\label{Lemma:poly_g(v=n-t)}
The error interpolation polynomial $\Tilde{g}(x)$ has the exact same form as the polynomial $a(x)$ in Theorem~\ref{Th:main} where $v=n-t$. 
\end{lemma}
\begin{IEEEproof}
The error interpolation polynomial $\Tilde{g}(x)$ with degree $m$ up to $n-1$ is evaluated on nonzero points in $\mathcal{A}$ and generates an error vector with $t$ nonzero components and $n-t$ zero components. Consequently, $\Tilde{g}(x)$ has $n-t$ nonzero distinct roots associated with the zero components of $\bm e$  due to the fact that our evaluation points in $\mathcal{A}$ are all nonzero. If the degree of $\Tilde{g}(x)$ is exactly $n-t$, then $u=0$ and $\Tilde{g}(x)$ satisfies (i) in Theorem~\ref{Th:main}. If $\deg(\Tilde{g}(x))=\lambda>n-t$, then it has an additional irreducible factor(s) besides its $n-t$ distinct linear factors. If the additional factors are all linear ($\lambda-n+t$ additional linear factors), then $\Tilde{g}(x)$ satisfies either (i) or (ii). If the additional factors are not all linear, then $\Tilde{g}(x)$ satisfies (iii). So our error interpolation polynomial $\Tilde{g}(x)$ has the exact same properties as we have for the polynomial $a(x)$ in Theorem~\ref{Th:main}. Hence, $g_i=a_i$ for $0\leq i\leq m-1$.
\end{IEEEproof}

\begin{lemma}\label{Th:nonsingular_toeplitz/circulant}
Let $\Tilde{\mat{G}}$ be the circulant matrix associated with the error interpolation polynomial $\Tilde{g}(x)$ where the weight of the error vector $\bm{e}$ is $t$. Then, the $t\times t$ Toeplitz submatrix formed by the first $t$ rows and the last $t$ columns of $\tilde{\mat{G}}$ is nonsingular. 
\end{lemma}
\begin{IEEEproof}
Using Lemma~\ref{Lemma:poly_g(v=n-t)}, if $v=n-t$ and $m\leq n-1$, then $g_i=a_i$ for $0\leq i \leq m-1$. Comparing the $t\times t$ Toeplitz matrix formed by the first $t$ rows and the last $t$ columns of $\tilde{\mat{G}}$ (colored in red) in \eqref{eq:tildeG} with the Toeplitz submatrix formed by the last $q-1-v$ rows and the last $q-1-v$ columns  of  $\mat{S}_v(x^{q-1}-1,a)$ (colored in red) in \eqref{eq:Sv} shows that these two submatrices are equal ($q-1-v=n-(n-t)=t$). It is proven in Theorem~\ref{Th:main} that the determinant of $\mat{S}_v(x^{q-1}-1,a)$  is nonzero and so the determinant of its $(q-1-v)\times (q-1-v)$ minor (colored in red) obtained by removing the first $m-v$ rows and columns, which dedicates the determinant of $\mat{S}_v(x^{q-1}-1,a)$, is nonzero. Thus, the determinant of the aforementioned $t\times t$ Toeplitz submatrix of $\tilde{\mat{G}}$ is nonzero, and therefore it is nonsingular. 
\end{IEEEproof}

\begin{continuance}{\ref{example1}}
   We have    
    $$\mat{S}_{7}(x^{10}-1,\Tilde{g})=  \small\left(\begin{array}{cc|ccc}
     1&~&2&~&~\\
     0&1&4&2&~\\
     0&0&\textcolor{red}{7}&\textcolor{red}{4}&\textcolor{red}{2}\\
     0&0&\textcolor{red}{8}&\textcolor{red}{7}&\textcolor{red}{4}\\
     0&0&\textcolor{red}{1}&\textcolor{red}{8}&\textcolor{red}{7}
    \end{array}\right).$$
    \normalsize
    Note that the two $3 \times 3$ submatrices colored in red  in $\tilde{\mat{G}}$ in \eqref{eq:tildeGexample} and $\mat{S}_{7}(x^{10}-1,\Tilde{g})$ are equal.  
    From Theorem~\ref{Th:main}, the determinant of $\mat{S}_7$ is  nonzero, and so is the determinant  of the  
  submatrix of $\tilde{\mat{G}}$ colored in red. The matrix $\tilde{\mat{G}}$ has rank $t=3$ and comparing with $\mat{S}_7$ and using Lemma~\ref{Th:nonsingular_toeplitz/circulant}, it follows that the $3\times 3$ submatrix in the top right corner of $\tilde{\mat{G}}$ is nonsingular.
\end{continuance}

\subsubsection{Reconstruction of $\Tilde{g}$}
As mentioned before, all the entries in the $t\times (t+1)$ submatrix located in the top right corner of $\tilde{\mat{G}}$ are known. Let $\tilde{\bm{g}}_j$ denote the $j$-th column of $\tilde{\mat{G}}$ for $0\leq i\leq n-1$. Using Lemma~\ref{Th:nonsingular_toeplitz/circulant}, we know that the column vectors $\Tilde{\bm{g}}_{n-t},\ldots, \tilde{\bm{g}}_{n-1}$ are independent and based on Theorem \ref{Th:konig-rados} the rank of $\tilde{\mat{G}}$ is $t$, hence we can write the first $t$ entries of $\tilde{\bm{g}}_{n-t-1}$ (colored in blue) as a linear combination of the first $t$ entries of the column vectors $\tilde{\bm{g}}_{n-t},\ldots, \tilde{\bm{g}}_{n-1}$ (colored in red). This gives the Toeplitz linear system of equations
\begin{equation*}
  {\small
  \textcolor{blue}{g_i}=\eta_1\textcolor{red}{g_{i+1}}+\eta_2\textcolor{red}{g_{i+2}}+\cdots+\eta_t\textcolor{red}{g_{i+t}},}
\end{equation*}
for $n-2t\leq i\leq n-t-1$, with $t$ equations and $t$ unknowns $\eta_1,\ldots, \eta_t$. The coefficient matrix for our system is nonsingular and a unique solution for $\eta_1,\ldots, \eta_t$ exists.%
\begin{continuance}{\ref{example1}}
  One can write the entries colored in blue as a linear combination of the entries colored in red,  resulting in the linear system of equations
  \begin{align*}
    \small
    \textcolor{red}{7}\eta_1+\textcolor{red}{4}\eta_2+\textcolor{red}{2}\eta_3&=\textcolor{blue}{8}\\
    \textcolor{red}{8}\eta_1+\textcolor{red}{7}\eta_2+\textcolor{red}{4}\eta_3&=\textcolor{blue}{1}\\
    \textcolor{red}{1}\eta_1+\textcolor{red}{8}\eta_2+\textcolor{red}{7}\eta_3&=\textcolor{blue}{2}.
  \end{align*}
  Since the coefficient matrix has a Toeplitz structure, we can use the \textnormal{\texttt{ADT/MDT}} algorithm in \cite{brent1980fast} to find its solution  $(\eta_1,\eta_2,\eta_3)=(6,1,3)$. 
\end{continuance}

Now, we can use the $\eta_i$'s and write the entries  $g_{k-1},\ldots, g_0$ of $\tilde{\bm{g}}_{n-t-1}$ as a linear combination of the column vectors $\tilde{\bm{g}}_{n-t},\ldots, \tilde{\bm{g}}_{n-1}$ in a Toeplitz form as
\begin{equation*}
  \small
  g_i=\eta_1g_{i+1}+\eta_2g_{i+2}+\cdots+\eta_tg_{i+t},
\end{equation*}
 for $0\leq i \leq n-2t-1$, which  recursively gives the unknown coefficients $g_{k-1},\ldots,g_0$. When the polynomial $\Tilde{g}(x)$ is found, we can compute $\bm{e}=\tilde{\bm{g}}\cdot \mat{G}_{\alpha}$ to find the error vector, and finally subtract $\bm e$ from the received word $\bm r$ gives the codeword $\bm{c}$. 
\begin{continuance}{\ref{example1}}
     In this step, we work on the submatrix obtained by removing the first $n-t-1=6$ columns, the first $t=3$ rows, and the last $t=3$ rows of $\tilde{\mat{G}}$. We use the coefficients $(\eta_1,\eta_2,\eta_3)=(6,1,3)$ computed above to find $g_3$ first, and then $g_2,g_1,g_0$, recursively, as
\begin{align*}
\small
  g_3&=2\eta_1+1\eta_2+8\eta_3\\
  g_2&=g_3\eta_1+2\eta_2+1\eta_3\\
  g_1&=g_2\eta_1+g_3\eta_2+2\eta_3\\
  g_0&=g_1\eta_1+g_2\eta_2+g_3\eta_3
\end{align*}
using methods in \cite{brent1980fast} and obtain $(g_3,g_2,g_1,g_0)=(4,7,8,1)$.
 Next, we compute  the error vector $\bm{e}=\tilde{\bm{g}}\cdot \mat{G}_{\alpha}=(0, 0, 0, 0, 5, 0, 4, 0, 1, 0 )$, and finally the corrected codeword $\bm{c}=\bm{r}-\bm{e}=( 8, 0, 4, 3, 1, 10, 8, 8, 3, 3 )\in \mathcal{C}$.
 \end{continuance}
 Our decoding algorithm consists of the following four steps.
\begin{itemize}[leftmargin=1.20cm]
    \item[Step 1.] Compute $\bm{\beta}=\bm{r}\cdot \mat{G}_{\alpha}^{-1}$.
    \item[Step 2.] Solve the $t\times t$ Toeplitz linear system 
      \begin{equation*}
        \mat{T}\cdot \bm{\eta}=\small
        \left(\begin{array}{ccc}
        g_{n-t}&\ldots&g_{n-1}  \\
        g_{n-t-1}&\ldots&g_{n-2}\\
        \vdots& \ddots&\vdots\\
        g_{n-2t+1}&\ldots&g_{n-t}
    \end{array}\right)\cdot \left(\begin{array}{c}
    \eta_1\\
    \eta_2\\
    \vdots\\
    \eta_{t}
    \end{array}\right)=\left(\begin{array}{c}
    g_{n-t-1}\\
    g_{n-t-2}\\
    \vdots\\
    g_{n-2t}
    \end{array}\right),
    \end{equation*}
 and find $\bm{\eta}=(\eta_1,\ldots,\eta_{t})$. The column vector in the right hand side contains the first $t$ entries of  $\tilde{\bm{g}}_{n-t-1}$.
  \item[Step 3.] Use $\bm{\eta}$ from Step 2 to solve the $k\times k$ Toeplitz linear system
    \begin{equation*}
    \hat{\mat{T}}\cdot \bm{\eta} = \small
    \left(\begin{array}{ccc}
        g_{k}&\ldots&g_{k+t-1}  \\
        g_{k-1}&\ldots&g_{k+t-2}\\
        \vdots& \ddots&\vdots\\
        g_{1}&\ldots&g_{t}
    \end{array}\right)\cdot \left(\begin{array}{c}
    \eta_1\\
    \eta_2\\
    \vdots\\
    \eta_{t}
    \end{array}\right)=\left(\begin{array}{c}
    g_{k-1}\\
    g_{k-2}\\
    \vdots\\
    g_{0}
    \end{array}\right),
    \end{equation*}
    %
    and obtain $g_0,\ldots, g_{k-1}$.
 \item[Step 4.] Calculate the error values via $\bm{e}=\tilde{\bm{g}}\cdot \mat{G}_{\alpha}$. 
\end{itemize}
\normalsize
\section{Complexity Analysis}\label{sec:com-analysis}
We summarize our proposed decoding algorithm in \cref{Alg1} and relate the steps above to the line numbers in \cref{Alg1} in the following discussion.

The first (\cref{alg:step1a,alg:step1b}) and the fourth (\cref{alg:step4}) steps are computed using an IDFT and a DFT, respectively, and since the matrix $\mat{G}_{\alpha}$ is a Vandermonde matrix and $\alpha$ is a primitive $n$-th root of unity, one can use an IFFT and an FFT, respectively, to accomplish them. When $\Fq$ is FFT-friendly,  the Cooley-Tukey FFT \cite{cooley1965algorithm}, which requires $\mathcal{O}(n\log n)$ field operations, can be applied and when $\Fq$ is not FFT-friendly, one can use the algorithm in \cite[Cor.~10.8]{von2013modern} with complexity $\mathcal{O}(n\log^2n\log\log n)$.
The only condition that we set for the decoding algorithm is $n=q-1$ and this guarantees the existence of a primitive $n$-th root of unity in $\Fq$. 


Step 2 is the bottleneck of our decoding algorithm. In Steps 2 (\cref{alg:step2a,alg:step2b,alg:step2c}) and 3 (\cref{alg:step3a,alg:step3b,alg:step3c}) we solve two linear systems of equations with $t$ and $k$ variables,  respectively, where the coefficient matrix is a nonsingular Toeplitz matrix. In \cite{brent1980fast}, Brent \etal proposed an algorithm to solve Toeplitz linear systems, which again can involve an FFT. In particular, one can find the vector $\bm{\eta}$ in  \cref{alg:step2c} using the \texttt{ADT/MDT} algorithm by Brent \etal in two steps.  The first step is to find the inverse of $\mat{T}$ and the second step is to find $\bm{\eta}$ using the  \texttt{SOLVE} subalgorithm. For an FFT-friendly finite field $\Fq$, the complexity of the first step is of order $\mathcal{O}(t\log^2t)$, while the complexity of the second step is $\mathcal{O}(t\log t)$. For an arbitrary finite field, however, the overall complexity becomes $O(t\log^2 t\log \log t)$.
%
Step 3 is  a convolution, and one can compute $g_{k-1},\ldots, g_{0}$   (\cref{alg:step3c}) using an FFT  with complexity $\mathcal{O}(k\log k)$ over an FFT-friendly finite field or with complexity $\mathcal{O}(k\log k\log\log k)$ using a DFT over an arbitrary finite field \cite[Th.~8.23]{von2013modern},~\cite{schonhage1977schnelle}. More methods to solve Toeplitz linear equation systems can be found in \cite{brent1988old,martinsson2005fast}. 
%
%

In summary, the overall complexity becomes $\mathcal{O}(n\log n+t\log^2 t) = \mathcal{O}(t\log^2 t)$ (as $t$ is of order $\mathcal{O}(n)$, but smaller) and $\mathcal{O}(n\log^2 n \log\log n)$ over an FFT-friendly and arbitrary finite field, respectively. 
We summarize the complexity discussion   with the following theorem.  

\begin{theorem}\label{Th:main-theorem}
Consider an $(n,k)$ RS code over $\Fq$, obtained by evaluation on $\mathcal{A}=\{\alpha^0,\alpha^1,\ldots,\alpha^{n-1} \}$ where $\alpha$ is  a primitive $n$-th root of unity and $n=q-1$. Then, every received word can be uniquely decoded up to $t\leq \lfloor \frac{n-k}{2}\rfloor$ errors using \cref{Alg1} with asymptotic complexity $\mathcal{O}(t\log^2t)$ and $\mathcal{O}(n\log^2n\log\log n)$, respectively, for FFT-friendly and arbitrary finite fields $\mathbb{F}_q$.
\end{theorem}

\begin{algorithm}[t]
		\small
		\SetAlgoLined
		\KwIn{A received word $\bm{r}$ with $t\leq \lfloor \frac{n-k}{2}\rfloor$ errors and distinct evaluation points $\alpha_i = \alpha^i$}
		\KwOut{The corrected codeword $\bm{c}\in \mathbb{F}_q^n$}
		Calculate $\beta(x)=\sum_{i=0}^{n-1}\beta_i x^{i}$ such that $\beta(\alpha_i) = r_i$ \label{alg:step1a}\\ 
            $(g_{k},\ldots, g_{n-1})\gets (\beta_k,\ldots,\beta_{n-1})$ \label{alg:step1b}\\
            $\bm{b} \gets ( g_{n-t-1},g_{n-t-2},\ldots,g_{n-2t})$ \label{alg:step2a}\\
            $\mat{T} \gets \texttt{ToepMat}_{t\times t}(g_{n-2t+1},\ldots,g_{n-t},\ldots,g_{n-1})$ \label{alg:step2b}\\
		$\bm{\eta}=(\eta_1,\ldots,\eta_{t}) \leftarrow  \texttt{ADT/MDT}(\mat{T}, \bm{b})$  \label{alg:step2c} \\
            $\hat{\mat{T}}\gets \texttt{ToepMat}_{k\times t}(g_{1},\ldots,g_{k},\ldots,g_{k+t-1})$ \label{alg:step3a}\\
            $\hat{\bm{b}}\gets (g_{k-1},\ldots,g_{0})$ \label{alg:step3b}\\
            $(g_{k-1},\ldots,g_{0}) \leftarrow \texttt{FFT/DFT}(\hat{\mat{T}}, \bm{\eta}, \hat{\bm{b}})$ \label{alg:step3c}\\
            $\bm{c} \leftarrow \bm{r}-\tilde{\bm{g}}\cdot \mat{G}_{\alpha}$ \label{alg:step4}\\
		\caption{Interpolation-based decoding}
		\label{Alg1}
		\normalsize
	\end{algorithm}
 \vspace{-2ex}

%

 
\section{Conclusion}
We proposed a new interpolation-based decoding algorithm for $(n,k)$ RS codes over finite fields $\Fq$, where $n=q-1$. We showed that it can correct any $t\leq \lfloor\frac{n-k}{2}\rfloor$ errors with complexity $\mathcal{O}(t\log^2 t)$ and $\mathcal{O}(n\log^2 n \log\log n)$ for  FFT-friendly and arbitrary finite fields, respectively.  
It is based on properties of a circulant matrix associated with the error interpolation polynomial and some known results from elimination theory. 

\IEEEtriggeratref{18}

\bibliographystyle{IEEEtran}
\bibliography{ref.bib}










\end{document}